\documentstyle[seceq]{ptptex}

\notypesetlogo  

\title{
Fermion Mass Hierarchy in the Grand Unified Theory\\
on $S_1/(Z_2 \times Z_2')$ Orbifold
}

\author{
N. {\sc Haba$^{1,2}$}\footnote{haba@eken.phys.nagoya-u.ac.jp},
Y. {\sc Shimizu$^2$}\footnote{shimizu@eken.phys.nagoya-u.ac.jp},
Tomoharu {\sc Suzuki$^2$}\footnote{tomoharu@eken.phys.nagoya-u.ac.jp}\\
and 
Kazumasa {\sc Ukai$^2$}\footnote{ukai@eken.phys.nagoya-u.ac.jp}
}

\inst{
$^1$Faculty of Engineering, Mie University,
Tsu Mie 514-8507, Japan\\
$^2$Department of Physics, Nagoya University,
Nagoya, 464-8602, Japan
}

\abst{

We suggest a simple grand unified theory 
 where the fifth dimensional coordinate is 
 compactified on an $S_1/(Z_2 \times Z_2')$ orbifold. 
This model is based on the supersymmetric 
 flipped $SU(5) \times U(1)$
 grand unified theory, which 
 can realize not only the triplet-doublet splitting 
 but also the natural fermion mass hierarchies. 
The triplet-doublet splitting is realized 
 by $S_1/(Z_2 \times Z_2')$ orbifolding, which 
 also reduces the gauge group as 
 $SU(5) \times U(1) \rightarrow SU(3)_c \times 
 SU(2)_L \times U(1)_Z \times U(1)_X$. 
The suitable fermion mass hierarchies are 
 generated by integrating out 
 extra three sets of vector-like heavy fields 
 which can propagate in five dimensions. 
The radiative corrections to 
 can reduce the gauge group to $SU(3)_c \times 
 SU(2)_L \times U(1)_Y$ by a simple extension of
 the model. 
}

\begin{document}

\maketitle
\section{Introduction} \label{sec:intro}

In the grand unified theory (GUT), 
 one of the most serious problems is 
 how to realize the mass splitting 
 between the triplet and 
 the doublet Higgs particles 
 in the Higgs sector. 
This problem is so-called 
 triplet-doublet (TD) splitting problem. 
{}For solving this serious problem, 
 people have suggested various 
 solutions, for example, the missing partner 
 mechanism\cite{missing}\cite{Yanagida:1995vq}\cite{flipped}\cite{Dedes},
 the idea of Higgs doublets as pseudo Nambu-Goldstone 
 bosons\cite{pG}, 
 the Dimopoulos-Wilczek mechanism\cite{DW}, 
 sliding singlet mechanism\cite{sl}, 
 and so on.
Recently, the new idea for solving the TD splitting problem 
 has been suggested in five dimensional $SU(5)$ 
 GUT where the fifth dimensional 
 coordinate is compactified on an
 $S_1/(Z_2 \times Z_2')$ 
 orbifold\cite{kawamura}
\cite{HN}\cite{others}\cite{others2}. 
In this model, only 
 Higgs and gauge fields can propagate in
 five dimensions, and the TD splitting 
 is realized by the same origin as 
 the gauge group reduction.


In this paper, 
 we consider the supersymmetric 
 flipped $SU(5) \times U(1)$ GUT in five dimensions. 
The fifth dimensional coordinate is 
 compactified on 
an $S_1/(Z_2 \times Z_2')$ orbifold. 
The model we propose 
 can realize not only TD splitting 
 but also the natural fermion mass hierarchies. 
The TD splitting is realized 
 by $S_1/(Z_2 \times Z_2')$ orbifolding, which 
 also reduces the gauge group as 
 $SU(5) \times U(1) \rightarrow SU(3)_c \times 
 SU(2)_L \times U(1)_Z \times U(1)_X$. 
The $U(1)_R$ symmetry protects 
 the Higgs doublets from 
 gaining heavy masses. 
The higher order operators do not destroy 
 the TD splitting in this model. 
In addition to three generation chiral matter
 fields, we introduce extra three sets of vector-like  
 matter fields which can propagate in five dimensions.  
The suitable fermion mass hierarchies are 
 generated by integrating out 
 these extra vector-like heavy fields. 
The large (small) flavor mixings in the lepton 
 (quark) sector are naturally explained. 
The proton-decay process through the dimension
 five operator is strongly suppressed by 
 $U(1)_R$ symmetry\cite{HN}, and  
 the dominant proton-decay mode is 
 $p \rightarrow e^+ \pi^0$ via the 
 exchange of $X,Y$ gauge bosons with 
 Kaluza-Klein masses. 
In the model we propose, the radiative corrections of 
 the large Yukawa coupling of right-handed neutrinos 
 can reduce the gauge group as 
 $SU(3)_c \times 
 SU(2)_L \times U(1)_Z \times 
 U(1)_X \rightarrow SU(3)_c \times 
 SU(2)_L \times U(1)_Y$ by a simple extension of
 the model. 

In section 2, we show the field contents of 
 this model and the gauge group reduction. 
In section 3, we will see the mechanism of
 generating the fermion mass hierarchies. 
Section 4 gives summary and discussions.

\section{Flipped $SU(5) \times U(1)$ GUT on $S_1/(Z_2 \times Z_2')$}

We denote the five dimensional coordinate 
 as $y$, which is 
 compactified on an $S_1/(Z_2 \times Z_2')$ 
 orbifold\footnote{
The five dimensional SUSY standard model 
 compactified on an $S_1/(Z_2 \times Z_2')$ 
 orbifold has been constructed in Refs.\cite{CSM}. 
There had been several works of the SUSY reduction
 by the compactification, for example, 
 in Refs.\cite{cern}\cite{LBL}. 
The extensions of the discrete symmetry and the 
 gauge symmetry are also 
 discussed in Refs.\cite{kawamoto} and
 \cite{others3}, respectively. 
}. 
Under the parity transformation of $Z_2$  and $Z_2'$, 
 which transforms $y \rightarrow -y$ and 
 $y' \rightarrow -y'$ ($y'=y+ \pi R/2$), respectively, 
 a field $\phi(x^\mu,y)$ which can propagate in five dimensions 
transforms as 
\begin{eqnarray}
\phi(x^\mu,y) &\to& \phi(x^\mu,-y) = P\phi(x^\mu,y),\\
\phi(x^\mu,y') &\to& \phi(x^\mu,-y') = P'\phi(x^\mu,y'),
\end{eqnarray}
where $P$ and $P'$ are operators of 
 $Z_2$  and $Z_2'$ transformations, respectively. 
Two walls at $y=0$ $( \pi R)$ and $\pi R/2$ $(- \pi R/2)$ are
 fixed points under $Z_2$ and $Z_2'$ transformations, 
 respectively. 
The physical space can be taken to be $0 \leq y \leq \pi R/2$,  
 since the walls at $y = \pi R$ and $- \pi R/2$ are 
 identified with those at $y = 0$ and $\pi R/2$, respectively. 
On this orbifold, the field $\phi(x^\mu,y)$ 
 is divided into 
\begin{eqnarray}
\phi_{++}(x^\mu,y) &=& \sum_{n=0}^{\infty} 
 \frac{1}{\sqrt{2^{\delta_{n,0}}\pi R}}\phi_{++}^{(2n)}
          (x^\mu)\cos \frac{2ny}{R},\\
\phi_{+-}(x^\mu,y) &=& \sum_{n=0}^{\infty} 
  \frac{1}{\sqrt{\pi R}}\phi_{+-}^{(2n+1)}
          (x^\mu)\cos \frac{(2n+1)y}{R},\\
\phi_{-+}(x^\mu,y) &=& \sum_{n=0}^{\infty} 
  \frac{1}{\sqrt{\pi R}}\phi_{-+}^{(2n+1)}
          (x^\mu)\sin \frac{(2n+1)y}{R},\\
\phi_{--}(x^\mu,y) &=& \sum_{n=0}^{\infty} 
   \frac{1}{\sqrt{\pi R}}\phi_{--}^{(2n+2)}
         (x^\mu)\sin \frac{(2n+2)y}{R},
\end{eqnarray}
according to the eigenvalues $( \pm, \pm)$ 
 of the parity 
 $(Z_2, Z_2')$.

Now let us see the supersymmetric flipped 
 $SU(5) \times U(1)$ GUT, which produces the 
 natural fermion mass hierarchies. 
The fifth dimensional coordinate is 
 compactified on 
the $S_1/(Z_2 \times Z_2')$ orbifold. 
We introduce three sets of 
 extra vector-like matter fields
 which can propagate in the bulk.
We take the $Z_2$ parity operator as $P= diag.(1,1,1,1,1)$ and 
the $Z_2'$ parity operator as $P'= diag.(-1,-1,-1,1,1)$ 
acting on a ${\bf 5}$ representation in 
 $SU(5)$\cite{kawamura}
\cite{HN}\cite{others}.

At first let us show matter multiplets. 
The ordinal chiral matter fields 
 are given by 
\begin{eqnarray}
\label{miracle}
& & ({\bf 10}_{1})_i = ( Q_L, 
       \overline{D_R^c}, \overline{N_R^c} )_i \;, \nonumber \\
& & ({\overline{\bf 5}_{-3})_i} = ( \overline{U_R^c}, L_L  )_i \;, \\
& & ({\bf 1}_{5})_i = ( \overline{E_R^c} )_i \;, \nonumber
\end{eqnarray}
where the index number shows the charge of $U(1)$, and 
the generation is denoted by $i = 1, 2, 3$. 
We assume that these chiral matter fields can not propagate 
 in the bulk and are localized on the 
 four dimensional wall at $y=0$ $(\pi R)$. 
The gauge quantum numbers after the compactification 
 and $U(1)_R$ charges 
 of these fields are shown in Table \ref{tb:particles1}.
The superpotential of the Yukawa sector 
 on the wall $y=0$ $(\pi R)$ is given by
\begin{equation}
\label{W_Y}
 W_Y = 
 y^d_{ij} H_5 {\bf 10}_i {\bf 10}_j
+ y^u_{ij} H_{\overline{5}} {\bf \overline{5}}_i {\bf 10}_j
+ y^e_{ij} H_5 {\bf \overline{5}}_i {\bf 1}_j 
+ y^{\nu}_{ij} H_{\overline{10}} {\bf 10}_i \phi_j 
+ M_{ij}^\phi \phi_i \phi_j.
\end{equation}
The field 
$\phi_i$ with the symmetric mass matrix $M_{ij}^\phi$, 
 being the origin of lepton number violation, 
 is the gauge singlet matter field
 localized on the four dimensional wall at $y=0$ $(\pi R)$, 
 which plays a crucial role for 
 making neutrino masses be light 
 as will be shown in the next section. 
We assume the eigenvalues of $M_{ij}^\phi$ is much larger than 
 the electroweak scale. 
All Yukawa couplings in Eq.(\ref{W_Y}) are 
 assumed to be of $O(1)$ independently 
 of the generation index.
$H$s represent Higgs fields which can 
 propagate in the five dimensions. 

In addition to above three generation chiral matter
 fields, we introduce extra three sets of vector-like  
 matter fields which can propagate in the bulk.
As we will show later, the suitable fermion mass hierarchies are 
 generated by integrating out these extra vector-like heavy fields. 
The gauge quantum numbers after the compactification, 
 the charges of $U(1)_R$ symmetry, 
 parity eigenvalues of $Z_2 \times Z_2'$, 
 and mass spectra at the tree level 
 are shown in Table \ref{tb:particles2}, 
where the index $I = 4, 5, 6$ denotes 
 the label of three sets of vector-like matter fields. 
The Yukawa interactions which mix 
 the ordinal chiral matter fields and extra 
 vector-like matter
 fields on the wall $y=0$ $(\pi R)$ are given by 
\begin{eqnarray}
\label{WY'}
 W_{Y'} &=& 
  y^{A}_{Ij} H_5 {\bf 10}_I {\bf 10}_j 
+ y^{B}_{Ij} H_{\overline{5}} {\bf \overline{5}}_I {\bf 10}_j
+ y^{C}_{iJ} H_{\overline{5}} {\bf \overline{5}}_i {\bf 10}_J
+ y^{D}_{iJ} H_5 {\bf \overline{5}}_i {\bf 1}_J
+ y^{E}_{Ij} H_5 {\bf \overline{5}}_I {\bf 1}_j  \nonumber \\
& &
  + y^{F}_{Ij} H_{10} {\bf \overline{10}}_I \phi_j 
  + y^{G}_{Ij} H_{\overline{10}} {\bf 10}_I \phi_j
  + y^{K}_{IJ} H_5 {\bf 10}_I {\bf 10}_J   
  + y^{L}_{IJ} 
    H_{\overline{5}} {\bf \overline{10}}_I {\bf \overline{10}}_J \\
& &
  + y^{P}_{IJ} H_5 {\bf 5}_I {\bf \overline{10}}_J 
  + y^{Q}_{IJ} H_{\overline{5}} {\bf \overline{5}}_I {\bf 10}_J 
  + y^{R}_{IJ} H_5 {\bf \overline{5}}_I {\bf 1}_J 
  + y^{S}_{IJ} 
H_{\overline{5}} {\bf 5}_I {\bf \overline{1}}_J . \nonumber
\end{eqnarray}
%
The volume suppression of extra dimension suggests 
 $y^{A \sim S} \ll 1$. 
The extra matters can have gauge invariant 
 vector-like mass terms, 
\begin{equation}
\label{WY''}
 W_{M} = 
   M_{IJ} {\bf 10}_I {\bf \overline{10}}_J
  + M_{IJ}' {\bf 5}_I {\bf \overline{5}}_J
  + M_{IJ}'' {\bf 1}_I {\bf \overline{1}}_J 
  + m_{iI} {\bf 10}_i {\bf \overline{10}}_I
  + m_{iI}' {\bf \overline{5}}_i {\bf 5}_I 
  + m_{iI}'' {\bf 1}_i {\bf \overline{1}}_I 
\end{equation}
on $y=0$ $( \pi R)$. 
{}For simplicity, we take 
 $M_{IJ} = M_{IJ}' = M_{IJ}'' = M_I \delta_{IJ}$  
 and $m_{iJ} = m_{iJ}' = m_{iJ}'' = m_i \delta_{i(J-3)}$. 
$M_{I}$ and $m_{i}$ include the 
 volume suppression factors of extra dimension, and 
 their values are assumed to be smaller
 than the compactification scale of $M_c ( \sim 1/R)$ 
 but much larger than the SUSY breaking scale 
 to avoid the blow-up of the gauge coupling constants. 
The Kaluza-Klein zero 
 modes of vector-like matter fields 
 obtain supersymmetric mass terms of $M_I$ and $m_i$.  
The ratios of $M_I$ and $m_i$ play
 crucial roles for generating fermion mass
 hierarchies as will be seen in the next section.

Table \ref{tb:particles3} shows the gauge quantum numbers 
 after the compactification, 
 the charges of $U(1)_R$ symmetry,
 parity eigenvalues of $Z_2 \times Z_2'$, 
 and mass spectra at the tree level 
 of Higgs super-multiplets. 
From this table it is found that the TD splitting is realized
 automatically  by the compactification, 
 since the doublet (triplet) Higgs fields $H_W$ and 
 $H_{\overline{W}}$ ($H_C$ and 
 $H_{\overline{C}}$)
 are (not) containing the Kaluza-Klein zero mode. 
This is the great benefit of the orbifold compactification 
 of $S_1/(Z_2 \times Z_2')$\cite{kawamura}\cite{HN}\cite{others}.

The superpotential of the Higgs sector 
 on the four dimensional wall at
 $y=0$ $(\pi R)$ is 
%
$W_H = 0 $
%
%
%
due to the $U(1)_R$ symmetry. 
We should notice that
 the conventional flipped $SU(5) \times U(1)$ GUT 
 in four dimensions have the Higgs superpotential as 
 $W_H^{(4d)} \simeq  H_{10} H_{10} H_{5} + 
 H_{\overline{10}} H_{\overline{10}}H_{\overline{5}}$, 
 which are needed for TD splitting. 
However, these Higgs interactions are not required 
 in our five-dimensional theory, since the 
 TD splitting is already realized by
 the orbifolding. 
As for Higgs mass parameters, 
 $\mu$ term, $\mu H_{5} H_{\overline{5}}$, and 
 soft SUSY breaking mass term, 
 $B \mu {h}_{5} {h}_{\overline{5}}$, 
 where ${h}_i$s denote 
 the scalar components of Higgs superfields, 
 are assumed to be generated 
 after the supersymmetry is broken. 
On the other hand, 
 $\mu ' H_{10}H_{\overline{10}}$  
 and $B' \mu ' {h}_{10} {h}_{\overline{10}}$ terms 
 must be forbidden by introducing 
 $Z_3$ $R$ 
 symmetry\cite{flipped}\cite{Dedes}\footnote{The higher 
 order superpotential interaction 
 $(H_{10}H_{\overline{10}})^n$ $(n \geq 2)$ is 
 also forbidden by this symmetry. },  
 unless $H_{10}$ and $H_{\overline{10}}$ do not take 
 large vacuum expectation values (VEVs) which 
 are needed for inducing the light neutrino masses.

Let us consider the supersymmetry breaking mechanism.
We introduce a gauge and $U(1)_R$ singlet field 
 $S = F_S \theta^2$ on the 
 $y= \pm \pi R/2$ branes\cite{HN}. 
The bulk gauge fields can 
 couple to the SUSY breaking 
 field $S$, and generate gaugino masses and   
the sfermion soft breaking masses are 
 induced through radiative corrections on 
 the four dimensional wall at $y=0 (\pi R)$. 
This SUSY breaking scenario is so-called gaugino mediation mechanism
\cite{gm}.
$\mu H_{5} H_{\overline{5}}$ and 
 $B \mu {h}_{5} {h}_{\overline{5}}$ terms 
 are also induced from the interactions 
 between bulk Higgs fields and $S$.

Next let us study the gauge group reduction. 
The gauge group is reduced by the $Z_2'$ parity 
 as $SU(5) \times U(1) \rightarrow SU(3)_c \times 
 SU(2)_L \times U(1)_Z \times U(1)_X$ at the 
 compactification scale 
 $M_c$. 
The VEVs 
 of $\langle h_N \rangle$ and/or 
 $\langle h_{\overline{N}}\rangle$
 reduce the gauge group as $SU(3)_c \times 
 SU(2)_L \times U(1)_Z \times U(1)_X \rightarrow SU(3)_c \times 
 SU(2)_L \times U(1)_Y$\footnote{In the 
 ordinary four-dimensional flipped 
 $SU(5) \times U(1)$ theory, 
 the VEVs of $h_{10}$ and 
 $h_{\overline{10}}$ can be 
 always identified with those of $h_{N}$ and 
 $h_{\overline N}$, respectively, by the field 
 rotation. 
This field redefinition is not available 
 in our five-dimensional theory. 
}. 
It is worth noting that 
 our model 
 do not demand the VEVs of $h_N$ and $h_{\overline{N}}$ 
 to be the GUT scale 
 contrary to the ordinary four-dimensional 
 flipped $SU(5) \times U(1)$ GUT. 
Are the components of 
 $h_N$ and $h_{\overline{N}}$ 
 really take the VEVs in 
 $h_{10}$ and $h_{\overline{10}}$ in our model
\footnote{The most simple way to 
 obtain the VEVs of $h_N$ and $h_{\overline{N}}$ 
 is setting the four-dimensional 
 superpotential $W_{v_{N}} \sim 
 X(H_N H_{\overline{N}} - v_{N}^2)$ 
 at $y= \pm \pi R/2$ branes where 
 there are no $SU(5) \times U(1)$ gauge transformations.
 Here $X$ is the gauge singlet field 
 localized on the $y= \pm \pi R/2$ branes.
}? 
Apparently, $h_Q$ and $h_{\overline{Q}}$ components 
 are too heavy to take 
 VEVs, since they have Kaluza-Klein 
 masses. 
Thus, in order to realize the suitable 
 gauge reduction, that means 
 $h_N$ and/or $h_{\overline{N}}$ take VEVs 
 whereas $h_D$ and/or $h_{\overline{D}}$ 
 do not, 
 the field contents in this model 
 need a little extension. 
We show a simple example here. 
We introduce vector-like matter fields 
 ${\bf 5'}_{-2}=(D', \overline{L}')$,  
 ${\bf \overline{5}'}_{2}=(\overline{D}', L')$, 
 ${\bf 10''}_{1}=(Q'', \overline{D}'', \overline{N}'')$, and 
 ${\bf \overline{10}''}_{-1}=(\overline{Q}'', D'', N'')$, which 
 are localized on the four dimensional wall at $y=0$ $(\pi R)$. 
Introducing an additional discrete symmetry, $P_2$, we assume 
 only these fields posses $P_2$ odd parity. 
Then, they have Yukawa couplings, 
\begin{equation}
\label{W5'}
W'= f H_{10}{\bf 10''}\ {\bf 5'}+ 
 \bar{f} H_{\overline{10}}
 {\bf \overline{10}''}\ {\bf \overline{5}'}+
 M_n  {\bf 5'}{\bf \overline{5}'} +
 M_m  {\bf 10''}{\bf \overline{10}''},
\end{equation}
where there are no mixings with other matter 
 fields due to $P_2$ symmetry. 
The values of $M_n$ and $M_m$ are smaller 
 than the compactification scale 
 but much larger than the SUSY breaking scale 
 to avoid the blow-up of the gauge coupling constants. 
We can evaluate the relevant soft SUSY 
 breaking masses using the following renormalization
 group equations (RGEs): 
\begin{eqnarray}
\label{e1}
\frac{dm^2_{{h}_{ D}}}{dt}
&=&
\frac{16}{3}g_3^2 M_3^2 + \frac{16}{9}g_Z^2 M_Z^2 +
 4g_X^{2}M_X^{2}
-{\rm tr}(y^\nu y^{\nu\dagger})m^2_{{h}_{ D}}\\
&&-{\rm tr}(y^\nu m^2_{\widetilde \phi}y^{\nu\dagger}
+ y^{\nu\dagger} m^2_{\widetilde {\overline D}} y^\nu )
-\left| \bar{f} \right|^2(m^2_{h_D}+m^2_{N''}+m^2_{\overline{D}'})
, \nonumber
\\
\label{e2}
\frac{dm^2_{{h}_{ N}}}{dt}
&=&
4g_Z^2M_Z^2 +
4g_X^{2}M_X^{2}
-{\rm tr}(y^\nu y^{\nu\dagger})m^2_{{h}_{ N}}
-{\rm tr}(y^\nu m^2_{\widetilde \phi}y^{\nu\dagger}
+ y^{\nu\dagger} m^2_{\widetilde {\overline N}} y^\nu )\nonumber \\
& & -3\left|\bar{f} \right|^2(m^2_{h_N}+m^2_{D''}+m^2_{\overline{D}'})
,
~~~~~~~\\
\label{e3}
\frac{dm^2_{{h}_{\overline{N}}}}{dt}
&=&
4g_Z^2M_Z^2 +
4g_X^{2}M_X^{2}
-3\left| f \right|^2(m^2_{h_{\overline{N}}}
 +m^2_{\overline{D}''}+m^2_{D'})
,
~~~~~~~\\
\label{e4}
\frac{d(m^2_{{\widetilde \phi}})_{ij}}{dt}
&=&
-2(m^2_{{\widetilde \phi}} y^{\nu\dagger} y^\nu 
 + y^{\nu\dagger} y^\nu m^2_{{\widetilde \phi}})_{ij}
- (y^{\nu\dagger} y^\nu)_{ij}m^2_{{h}_{ N}}
- (y^{\nu\dagger} m^2_{\widetilde {\overline N}} y^\nu)_{ij}
\nonumber\\
&&- 3(y^{\nu\dagger} y^\nu)_{ij}m^2_{{h}_{ D}}
- 3(y^{\nu\dagger} m^2_{\widetilde {\overline D}} y^\nu)_{ij} \ .
\end{eqnarray}
Here $t=-1/(4\pi)^2\ln(\mu^2)$. 
The above RGEs are available in the energy scale of  
 $M_{ij}^\phi < \mu < M_c$. 
Here we neglect the small Yukawa couplings 
 between matter fields 
 and extra generations 
 in Eq.(\ref{WY'})\footnote{The small couplings of 
 $y^{A \sim E}$ evade the large flavor changing neutral 
 current (FCNC) induced from the destruction of 
 sfermion mass degeneracies.
}. 
$g_3$, $g_Z$, and $g_X$ are
 gauge coupling constants for $SU(3)_c$, $U(1)_Z$, 
 and $U(1)_X$, respectively.
The RGEs show that  
 gauge couplings give the positive contributions 
 whereas the Yukawa couplings give the negative contributions
 to the soft breaking masses 
 toward the low energy scale. 
Then, in the case that the Yukawa coupling $y^\nu$ 
 is sufficiently large, scalar squared masses 
 can become negative through the 
 radiative corrections\cite{IKKT}. 
$h_N$ and $h_{\overline N}$ have 
 the $D$-flat 
 direction, $| \langle h_N \rangle | =
 | \langle h_{\overline{N}}\rangle | (\equiv v_N)$, 
 in the SUSY limit. 
The soft SUSY breaking terms $m^2_{{h}_{ N}}$ 
 and $m^2_{{h}_{\overline{N}}}$ induce a small
 deviation from flatness, and 
 the Higgs ``effective potential'' \cite{Dedes}
 is given by 
 $V \simeq [m^2_{{h}_{N}}(v_N)+
 m^2_{{h}_{\overline{N}}}(v_N) ] v_N^2$ 
 where $m^2( \mu )$ shows the value of RGE 
 running parameter $m^2$ at 
 energy scale $\mu$. 
Starting from high energy with positive soft 
 masses squared for $m^2_{{h}_{ N}}$ 
 and $m^2_{{h}_{\overline{N}}}$, 
 the RGE effects make a reversal of sign 
 for $m^2_{{h}_{ N}}+m^2_{{h}_{\overline{N}}}$. 
This reversal of sign shows the development 
 of the symmetry breaking minimum along the 
 flat direction with VEV of order $v_N$. 
We can check the large magnitudes of $f$ and $\bar{f}$ really 
 generate the large value of $v_N$\footnote{Equation 
 (\ref{e2}) shows that the large magnitude 
 of $y^{\nu}$ might induce negative contributions 
 to the RGEs of soft squared masses. 
 However, the running parameter 
 $m^2_{{h}_{ N}}+m^2_{{h}_{\overline{N}}}$ 
 does not become negative since the value of 
 $m^2_{\widetilde {\overline N}}$, 
 existing in the 3rd and 4th terms in the R.H.S. of 
 Eq.(\ref{e2}), 
 also become small by  the RGE effects. 
 That is why we introduce ${\bf 5'}, {\bf \overline{5}'}, 
 {\bf 10''}$, and ${\bf \overline{10}''}$.  
}. 
The value of $m^2_{{h}_{ N}}+m^2_{{h}_{\overline{N}}}$ 
 becomes negative faster than 
 that of $m^2_{{h}_{ D}}+m^2_{{h}_{\overline{D}}}$, 
 since $m^2_{{h}_{D}}$ and $m^2_{{h}_{\overline{D}}}$ receive 
 positive corrections from the QCD effects 
 (see Eqs.(\ref{e1})$\sim$(\ref{e3})).
Therefore we can conclude that the suitable gauge reduction 
 is really realized in this 
 model\footnote{When we consider the gravity mediated 
 SUSY breaking scenario, the large value of soft SUSY breaking 
 $A$ parameters might realize radiative 
 symmetry breaking of $SU(5) \times U(1)$ 
 as shown in Ref.\cite{Dedes}, 
 by using the $A$ term contribution appearing in 
 Eqs.(\ref{e2}) and (\ref{e3}).  
However, our model can not find this solution under the
 color and charge conserving sufficient 
 condition, $A < 3 m_{soft}$. 
}. 
As for 
 the physical squared mass of gauge singlet 
 field ${\widetilde \phi}_i$, it is 
 positive due to the 
 large supersymmetric masses of 
 $M_{ij}^{\phi}$ in Eq.(\ref{W_Y}), 
 even if the large value of $y^\nu$ causes the 
 negative contributions to 
 $m^2_{\widetilde \phi}$ in Eq.(\ref{e4}).


Here we should estimate the 
 corrections of higher order
 operators induced from the 
 VEVs of $\langle h_N \rangle \simeq 
 \langle h_{\overline{N}}\rangle$. 
The higher order operators  
 $\{ \langle h_N \rangle \langle h_{\overline{N}} 
 \rangle /M_c \} ({\bf 5} {\bf \overline{5}})$ 
 induce the corrections to the mass parameters 
 in Eq.(\ref{WY''}). 
In order not to destroy 
 the fermion mass hierarchies
 which will be shown in the next section, 
 the relation of 
 $\langle h_N \rangle \langle h_{\overline{N}}\rangle 
 /M_c \ll M_I, m_i$ should be satisfied. 
This constraint is satisfied when 
 the magnitudes of $\langle h_N \rangle$ and 
 $\langle h_{\overline{N}}\rangle$ are smaller 
 than $M_I$ and $m_i$.  
Under this condition, the corrections of 
 $\mu$ parameters, which is $( \langle h_N \rangle 
 \langle h_{\overline{N}}\rangle /M_c^2)
 \mu H_W H_{\overline{W}}$,  
 is negligibly small. 
Thus the higher order operator does not destroy 
 the TD splitting in this model.

As for the proton stability, 
 the $p \rightarrow e^+ \pi^0$ process 
 is dominant via 
 $X, Y$ gauge boson exchange, 
 which has mass of order $M_c ( \sim 1/R)$ 
 as shown in Table IV.  
The proton-decay process through the dimension
 five operator is strongly suppressed by 
 $U(1)_R$ symmetry\cite{HN}. 
It is because the colored Higgs $H_C$  
 $(H_{\overline{C}})$ has Kaluza-Klein mass of order 
 $M_c$ with $H^c_C$ $(H^c_{\overline{C}})$, and 
 the conjugate fields $H^c_C$ and $H^c_{\overline{C}}$ 
 do not couple directly to the quarks and leptons. 

\section{Fermion Mass Hierarchy}

Let us see the mechanism 
 which produces the fermion mass hierarchies
 in this model. 
The fermion mass hierarchies 
 in the chiral matter fields are generated by 
 integrating out the heavy extra vector-like 
 generations\cite{BB}. 
Let us see, for example,
 the quark doublet $(Q_i)$ sector. 
The mass terms of the quark doublet sector  
 in Eq.(\ref{WY''}) are given by  
\begin{equation}
 W_M = \sum_{i=1}^3 (m_i Q_{i} \overline{Q_{i+3}} + 
                     M_{i+3} Q_{i+3} \overline{Q_{i+3}}). 
\end{equation}
All these fields represent Kaluza-Klein zero mode. 
Then the light eigenstate $Q_i^{l}$, which 
 is just the quark doublet at the 
 low energy, and the heavy eigenstate $Q_i^{H}$ 
 are given by 
\begin{eqnarray}
\label{Q}
 Q_i^{l} &=&  {M_{i+3} \over \sqrt{M_{i+3}^2 + m_i^2}}Q_{i}
         -{m_i \over \sqrt{M_{i+3}^2 + m_i^2}}Q_{i+3}, \\
 Q_i^{H} &=&  {m_i \over \sqrt{M_{i+3}^2 + m_i^2}}Q_{i}+
          {M_{i+3} \over \sqrt{M_{i+3}^2 + m_i^2}}Q_{i+3}.  
\end{eqnarray}
We consider the case of 
 $\epsilon_1 \simeq M_4/m_1 \ll 1$,
 $\epsilon_2 \simeq M_5/m_2  < 1 $, and 
 $M_6/m_3 \sim 1$, where 
 $\epsilon_i \equiv {M_{i+3} / 
 \sqrt{M_{i+3}^2 + m_i^2}}$. 
Then, the mass hierarchy is generated 
 in the mass matrix of the 
 light eigenstate $Q_i^{l}$. 
The fields $\overline{U_i}$ and $\overline{E_i}$ 
 also receive the same effects as Eq.(\ref{Q}) in 
 the light eigenstates, 
 but $\overline{D_i}$, $L_i$, and
 $\overline{N_i}$ do not receive these effects
 since their extra vector-like generations 
 do not have zero modes as shown in Table \ref{tb:particles2}.

Bellow the electroweak scale, 
 the light eigenstates mass matrices of 
 up quark sector, down quark sector, 
 and charged lepton sector are 
 given by  
\begin{equation}
 m_u^l \simeq \left(
\begin{array}{ccc}
 \epsilon_1^2 & \epsilon_2 \epsilon_1 &  \epsilon_1  \\
 \epsilon_1 \epsilon_2 & \epsilon_2^2  &  \epsilon_2  \\
 \epsilon_1  & \epsilon_2  & 1
\end{array}
\right)  \overline{v}, \;\;
 m_d^l \simeq \left(
\begin{array}{ccc}
 \epsilon_1 & \epsilon_1 &  \epsilon_1  \\
 \epsilon_2 & \epsilon_2  &  \epsilon_2  \\
  1 & 1 & 1 
\end{array}
\right) v, \;\;
 m_e^l \simeq \left(
\begin{array}{ccc}
 \epsilon_1 & \epsilon_2 & 1  \\
 \epsilon_1 & \epsilon_2  & 1 \\
 \epsilon_1 & \epsilon_2 & 1 
\end{array}
\right) v, \;\;
\label{mass}
\end{equation}
respectively, 
where $v \equiv \langle h_W \rangle$,   
 $\overline{v}\equiv \langle h_{\overline{W}} \rangle$. 
Each element is understood to be multiplied by 
 $O(1)$ coefficient. 
We write the mass matrices 
 that the left-handed fermions are to the left
 and the right-handed fermions are to the right. 
Setting the values of $M_{i+3}$ and $m_i$ as 
 $\epsilon_1 \simeq \lambda^4$ and 
 $\epsilon_2 \simeq \lambda^2$, 
 where $\lambda$ is the Cabbibo angle estimated as $0.2$, 
 we can obtain the suitable mass hierarchies. 
{}Moreover, 
 the small (large) flavor mixings 
 in the quark (lepton) sector are naturally 
 obtained\cite{BB}\cite{BB1}\cite{anarchy}.

The neutrino mass matrix is given by 
\begin{equation}
 m_\nu = 
\bordermatrix{
 & L_i & \overline{N_i} & \phi_i & N_I^{(2n+1)}  & N_I^{c(2n+1)} \cr
 & 0 & m_\nu^D &  0 & y^{C} {\overline{v}} & 0 \cr
 &  m_\nu^{DT} & 0 & y^\nu \langle h_N \rangle & m_i & 0 \cr
 &  0 & y^\nu{}^T \langle h_N \rangle & M_{ij}^\phi & 
  y^{F} \langle h_{\overline{N}}\rangle & 0 \cr
 & y^{C}{}^T {\overline{v}} & m_i & 
  y^{F}{}^T \langle h_{\overline{N}}\rangle & 0 & (2n+1)/R  \cr
 & 0 & 0 & 0 & (2n+1)/R & 0
}
\label{nu}
\end{equation}
where the neutrino Dirac mass matrix is given by 
\begin{equation}
 m_\nu^D \simeq \left(
\begin{array}{ccc}
1 & 1 & 1 \\
1 & 1 & 1 \\
1 & 1 & 1
\end{array}
\right)  \overline{v} ,
\end{equation}
because the fields $L_I$, $\overline{L_I}$, 
 $\overline{N_I}$, and $N_I$ do not have 
 the Kaluza-Klein zero mode. 
Each element of $m_\nu^D$ has $O(1)$ 
 coefficient. 
Neglecting the contributions from the super-heavy 
 Kaluza-Klein masses $(2n+1)/R$, 
 Eq.(\ref{nu}) induces 
 the mass matrix of three light neutrinos as 
%
\begin{equation}
\label{26}
 m_\nu^{(l)} \simeq
 {m_\nu^D m_\nu^D{}^T \over y^\nu{}^2 v_N^2/M^\phi}= 
\left(
\begin{array}{ccc}
 1 & 1 & 1 \\
 1 & 1 & 1 \\
 1 & 1 & 1 
\end{array}
\right)  {\overline{v}^2 M^\phi \over y^\nu{}^2 v_N^2}.
\end{equation}
%
$M^\phi$ in Eq.(\ref{26}) means the typical scale
 of $M_{ij}^\phi$. 
We can obtain the suitable mass scale for
 the neutrino oscillation experiments 
 by choosing the values of $M^\phi$ and $y^\nu v_N$. 
The suitable choice of $O(1)$ coefficients 
 in the mass matrix can derive the suitable 
 flavor mixings consistent with the neutrino
 oscillation experiments.

Above fermion mass matrices can give the 
 suitable mass hierarchies of 
 quarks and leptons\cite{BB}\cite{BB1}\cite{anarchy}. 
They also give us the natural explanation 
 why the flavor mixing in the quark sector 
 is small while the flavor mixing in the 
 lepton sector is large.
Since all components have undetermined 
 $O(1)$ coefficients, we have to determine
 the explicit values of $O(1)$ coefficients
 in order to predict the experimentally observable
 quantities. 



\section{Summary and Discussion}

In this paper,  we have proposed a supersymmetric 
 flipped $SU(5) \times U(1)$ GUT in five dimensions 
where the fifth dimensional coordinate is 
 compactified on an $S_1/(Z_2 \times Z_2')$ orbifold. 
We have shown that the model can realize not only the TD splitting 
but also the natural fermion mass hierarchies. 
The TD splitting is realized 
by the $S_1/(Z_2 \times Z_2')$ orbifolding, which 
also reduces the gauge group as 
 $SU(5) \times U(1) \rightarrow SU(3)_c \times 
 SU(2)_L \times U(1)_Z \times U(1)_X$. 
The triplet Higgs fields have the Kaluza-Klein masses
of order of $1/R$ whereas the $U(1)_R$ symmetry protects 
 the Higgs doublets from gaining heavy masses. 
The higher order operators do not destroy 
 the triplet-doublet splitting in this model. 
The proton-decay process through the dimension
five operator is strongly suppressed by 
the $U(1)_R$ symmetry, and  the dominant proton-decay mode is 
$p \rightarrow e^+ \pi^0$ via the exchange of the $X,Y$ gauge 
bosons which have Kaluza-Klein masses.

A simple extension of
 the model  
 can make the SUSY breaking squared mass 
 of $h_N$ and $h_{\overline N}$ be negative, 
 which reduces the gauge group as 
 $SU(3)_c \times 
 SU(2)_L \times U(1)_Z \times 
 U(1)_X \rightarrow SU(3)_c \times 
 SU(2)_L \times U(1)_Y$.

In addition to three generation chiral matter fields, 
 we have introduced extra three sets of vector-like  
 matter fields which can propagate in the bulk.
The suitable fermion mass hierarchies are 
 generated by integrating out 
 these extra vector-like heavy fields. 
Moreover, 
 the large (small) flavor mixings in the lepton 
 (quark) sector are naturally explained.

\section*{Acknowledgment}
We would like to thank Y. Kawamura, Y. Nomura, and 
 T. Kondo for useful discussions. 
Research of KU is supported in part by the Japan Society for 
 Promotion of Science under the Predoctoral Research Program. 
This work is supported in
 part by the Grant-in-Aid for Science
 Research, Ministry of Education, Science and Culture, Japan
 (No.12004276, No.12740146, No.13001292).


\newpage
\begin{table}
\begin{center}
 \begin{tabular}[tb]{|c|l|c|}
\hline
 4d matter fields & $(SU(3)_C, SU(2)_L, U(1)_Z, U(1)_X)$ 
 & $U(1)_R$  \\
 \hline\hline
 $Q_i$ &   
 $({\bf 3}, {\bf 2}, {\bf 1/6}, {\bf 1})$
 &1  \\
 $\overline{D}_i$ &    
 $({\bf \overline{3}}, {\bf 1}, {\bf -2/3}, {\bf 1})$
 &1  \\
 $\overline{N}_i$ &    
 $({\bf 1}, {\bf 1}, {\bf 1}, {\bf 1})$
 &1  \\
 \hline
 $\overline{U}_i$ &    
 $({\bf \overline{3}}, {\bf 1}, {\bf 1/3}, {\bf -3})$
 &1  \\
 $L_i$ &    
 $({\bf 1}, {\bf 2}, {\bf -1/2}, {\bf -3})$
 &1  \\
 \hline
 $\overline{E}_i$ &    
 $({\bf 1}, {\bf 1}, {\bf 0}, {\bf 5})$
 &1  \\
\hline
 $\phi_i$ &    
 $({\bf 1}, {\bf 1}, {\bf 0}, {\bf 0})$
 &1  \\
\hline
 \end{tabular}
\end{center}
\caption{The gauge quantum numbers after the compactification and 
 $U(1)_R$ charges of the chiral matter 
 fields confined on the wall at $y=0$ $(\pi R)$
 are shown.
 The index $i = 1, 2, 3$ denotes the generation. }
\label{tb:particles1}
\end{table}

\begin{table}
\begin{center}
 \begin{tabular}[tb]{|l|l|c|c|c|}
\hline
 extra matter fields & extra matter fields 
 & & & \\
 $(SU(5), U(1)_X)$ & $(SU(3)_C, SU(2)_L, U(1)_Z, U(1)_X)$ 
 &$U(1)_R$& $(Z_2, Z_2')$ &
 mass \\
 \hline\hline
 ${\bf 10}_I({\bf 10}, {\bf 1})$ &    
 $Q_I^{(2n)}({\bf 3}, {\bf 2}, {\bf 1/6}, {\bf 1})$
  &1& $(+, +)$ & $\frac{2n}{R}$ \\
 &    
 $\overline{D}_I^{(2n+1)}({\bf \overline{3}}, {\bf 1}, {\bf -2/3}, {\bf 1})$, 
 $\overline{N}_I^{(2n+1)}({\bf 1}, {\bf 1}, {\bf 1}, {\bf 1})$
 &1&  $(+, -)$ & $\frac{2n+1}{R}$ \\
 \hline
 ${\bf 10}_I^c({\bf \overline{10}}, {\bf -1})$ &    
 $Q_I^{c(2n+2)}({\bf \overline{3}}, {\bf 2}, {\bf -1/6}, {\bf -1})$
 &$1$& $(-, -)$ & $\frac{2n+2}{R}$ \\
 &    
 $\overline{D}_I^{c(2n+1)}({\bf 3}, {\bf 1}, {\bf 2/3}, {\bf -1})$,
 $\overline{N}_I^{c(2n+1)}({\bf 1}, {\bf 1}, {\bf -1}, {\bf -1})$
 &$1$&  $(-, +)$ & $\frac{2n+1}{R}$ \\
 \hline
 ${\bf \overline{10}}_I({\bf \overline{10}}, {\bf -1})$ &   
 $\overline{Q}_I^{(2n)}({\bf \overline{3}}, {\bf 2}, {\bf -1/6}, {\bf -1})$
 &1&  $(+, +)$ & $\frac{2n}{R}$ \\
 &   
 $D_I^{(2n+1)}({\bf 3}, {\bf 1} ,{\bf 2/3}, {\bf -1})$, 
 $N_I^{(2n+1)}({\bf 1}, {\bf 1}, {\bf -1}, {\bf -1})$
 &1&  $(+, -)$ & $\frac{2n+1}{R}$ \\
 \hline
 ${\bf \overline{10}}_I^c({\bf 10}, {\bf 1})$ &   
 $\overline{Q}_I^{c(2n+2)}({\bf 3}, {\bf 2}, {\bf 1/6}, {\bf 1})$
 &$1$&  $(-,-)$ & $\frac{2n+2}{R}$ \\
 &   
 $D_I^{c(2n+1)}({\bf \overline{3}}, {\bf 1}, {\bf -2/3}, {\bf 1})$, 
 $N_I^{c(2n+1)}({\bf 1}, {\bf 1}, {\bf 1}, {\bf 1})$
 &$1$&  $(-, +)$ & $\frac{2n+1}{R}$ \\
 \hline
 ${\bf \overline{5}}_I({\bf \overline{5}}, {\bf -3})$ &    
 $\overline{U}_I^{(2n)}({\bf \overline{3}}, {\bf 1}, {\bf 1/3}, {\bf -3})$
 &1&  $(+, +)$ & $\frac{2n}{R}$ \\
 &   
 $L_I^{(2n+1)}({\bf 1}, {\bf 2}, {\bf -1/2}, {\bf -3})$
 &1&  $(+, -)$ & $\frac{2n+1}{R}$ \\
 \hline
 ${\bf \overline{5}}_I^c({\bf 5}, {\bf 3})$ &    
 $\overline{U}_I^{c(2n+2)}({\bf 3}, {\bf 1}, {\bf -1/3}, {\bf 3})$
 &$1$&  $(-, -)$ & $\frac{2n+2}{R}$ \\
 &   
 $L_I^{c(2n+1)}({\bf 1}, {\bf 2}, {\bf 1/2}, {\bf 3})$
 &$1$&  $(-, +)$ & $\frac{2n+1}{R}$ \\
 \hline
 ${\bf 5}_I({\bf 5}, {\bf 3})$ &   
 $U_I^{(2n)}({\bf 3}, {\bf 1}, {\bf -1/3}, {\bf 3})$
 &1&  $(+, +)$ & $\frac{2n}{R}$ \\
 &   
 $\overline{L}_I^{(2n+1)}({\bf 1}, {\bf 2}, {\bf 1/2}, {\bf 3})$
 &1&  $(+, -)$ & $\frac{2n+1}{R}$ \\
 \hline
 ${\bf 5}_I^c({\bf \overline{5}}, {\bf -3})$ &   
 $U_I^{c(2n+2)}({\bf \overline{3}}, {\bf 1} , {\bf 1/3}, {\bf -3})$
 &$1$&  $(-, -)$ & $\frac{2n+2}{R}$ \\
 &   
 $\overline{L}_I^{c(2n+1)}({\bf 1}, {\bf 2}, {\bf -1/2}, {\bf -3})$
 &$1$&  $(-, +)$ & $\frac{2n+1}{R}$ \\
 \hline
 ${\bf 1}_I({\bf 1}, {\bf 5})$ &   
 $\overline{E}_I^{(2n)}({\bf 1}, {\bf 1}, {\bf 0}, {\bf 5})$
 &1&  $(+, +)$ & $\frac{2n}{R}$ \\
 \hline
 ${\bf 1}_I^c({\bf 1}, {\bf -5})$ &   
 $\overline{E}_I^{c(2n+2)}({\bf 1}, {\bf 1}, {\bf 0}, {\bf -5})$
 &$1$&  $(-, -)$ & $\frac{2n+2}{R}$ \\
 \hline
 ${\bf 1}_I({\bf 1}, {\bf -5})$ &   
 $E_I^{(2n)}({\bf 1}, {\bf 1}, {\bf 0}, {\bf -5})$
 &1&  $(+, +)$ & $\frac{2n}{R}$ \\
 \hline
 ${\bf 1}_I^c({\bf 1}, {\bf 5})$ &   
 $E_I^{c(2n+2)}({\bf 1}, {\bf 1}, {\bf 0}, {\bf 5})$
 &$1$&  $(-, -)$ & $\frac{2n+2}{R}$ \\
 \hline
 \end{tabular}
\end{center}
\caption{The gauge quantum numbers after the compactification, 
 the charges of $U(1)_R$ symmetry, 
 parity eigenvalues of $Z_2 \times Z_2'$, 
 and mass spectra at the tree level of the three sets of 
 vector-like extra matter fields 
 are shown.
 The index $I = 4, 5, 6$ denotes the label of extra matter fields.
 The first column represents the corresponding extra matter fields 
before the compactification.} 
\label{tb:particles2}
\end{table}

\begin{table}
\begin{center}
 \begin{tabular}[tb]{|l|l|c|c|c|}
\hline
 Higgs fields & Higgs fields & & & \\
 $(SU(5), U(1)_X)$  &  
 $(SU(3)_C, SU(2)_L, U(1)_Z, U(1)_X)$ 
 & $U(1)_R$ & $(Z_2, Z_2')$ &
 mass \\
 \hline\hline
 $H_{10}({\bf 10}, {\bf 1})$ &   
 $H_Q^{(2n+1)}({\bf 3}, {\bf 2}, {\bf 1/6}, {\bf 1})$
 &0  & $(+, -)$ & $\frac{2n+1}{R}$ \\
 &    
 $H_{\overline{D}}^{(2n)}({\bf \overline{3}}, {\bf 1}, {\bf -2/3}, {\bf 1})$,  
 $H_{\overline{N}}^{(2n)}({\bf 1}, {\bf 1}, {\bf 1}, {\bf 1})$
 &0 & $(+, +)$ & $\frac{2n}{R}$ \\
 \hline
 $H_{10}^c({\bf \overline{10}}, {\bf -1})$ &   
 $H_Q^{c(2n+1)}({\bf \overline{3}}, {\bf 2}, {\bf -1/6}, {\bf -1})$
 &2  & $(-, +)$ & $\frac{2n+1}{R}$ \\
 &    
 $H_{\overline{D}}^{c(2n+2)}({\bf 3}, {\bf 1}, {\bf 2/3}, {\bf -1})$,  
 $H_{\overline{N}}^{c(2n+2)}({\bf 1}, {\bf 1}, {\bf -1}, {\bf -1})$
 &2 & $(-, -)$ & $\frac{2n+2}{R}$ \\
 \hline
 $H_{\overline{10}}({\bf \overline{10}}, {\bf -1})$ &   
 $H_{\overline{Q}}^{(2n+1)}({\bf \overline{3}}, {\bf 2}, {\bf -1/6}, {\bf -1})$
 &0  & $(+, -)$ & $\frac{2n+1}{R}$ \\
 &   
 $H_D^{(2n)}({\bf 3}, {\bf 1}, {\bf 2/3}, {\bf -1})$
,  $H_N^{(2n)}({\bf 1}, {\bf 1}, {\bf -1}, {\bf -1})$
 &0  & $(+, +)$ & $\frac{2n}{R}$ \\
 \hline
 $H_{\overline{10}}^c({\bf 10}, {\bf 1})$ &   
 $H_{\overline{Q}}^{c(2n+1)}({\bf 3}, {\bf 2}, {\bf 1/6}, {\bf 1} )$
 &2  & $(-, +)$ & $\frac{2n+1}{R}$ \\
 &   
 $H_D^{c(2n+2)}({\bf \overline{3}}, {\bf 1}, {\bf -2/3}, {\bf 1})$
,  $H_N^{c(2n+2)}({\bf 1}, {\bf 1}, , {\bf 1}, {\bf 1})$
 &2  & $(-, -)$ & $\frac{2n+2}{R}$ \\
 \hline
 $H_5({\bf 5}, {\bf -2})$ & 
 $H_C^{(2n+1)}({\bf 3}, {\bf 1},  {\bf -1/3}, {\bf -2})$
 &0  & $(+, -)$ & $\frac{2n+1}{R}$ \\
 &   
 $H_W^{(2n)}({\bf 1}, {\bf 2}, {\bf 1/2}, {\bf -2})$
 &0  & $(+, +)$ & $\frac{2n}{R}$ \\
 \hline
 $H_5^c({\bf \overline{5}}, {\bf 2})$ & 
 $H_C^{c(2n+1)}({\bf \overline{3}}, {\bf 1}, {\bf 1/3}, {\bf 2})$
 &2  & $(-, +)$ & $\frac{2n+1}{R}$ \\
 &   
 $H_W^{c(2n+2)}({\bf 1}, {\bf 2}, {\bf -1/2}, {\bf 2})$
 &2  & $(-, -)$ & $\frac{2n+2}{R}$ \\
 \hline
 $H_{\overline{5}}({\bf \overline{5}}, {\bf 2})$ & 
 $H_{\overline{C}}^{(2n+1)}({\bf \overline{3}}, {\bf 1} , {\bf 1/3}, {\bf 2})$
 &0  & $(+, -)$ & $\frac{2n+1}{R}$ \\
 &   
 $H_{\overline{W}}^{(2n)}({\bf 1}, {\bf 2},  {\bf -1/2}, {\bf 2})$
 &0  & $(+, +)$ & $\frac{2n}{R}$ \\
 \hline
 $H_{\overline{5}}^c({\bf 5}, {\bf -2})$ & 
 $H_{\overline{C}}^{c(2n+1)}({\bf 3}, {\bf 1} , {\bf -1/3}, {\bf -2})$
 &2  & $(-, +)$ & $\frac{2n+1}{R}$ \\
 &   
 $H_{\overline{W}}^{c(2n+2)}({\bf 1}, {\bf 2},  {\bf 1/2}, {\bf -2})$
 &2  & $(-, -)$ & $\frac{2n+2}{R}$ \\
 \hline
 \end{tabular}
\end{center}
\caption{The gauge quantum numbers 
 after the compactification,  
 the charges of $U(1)_R$ symmetry, 
 parity eigenvalues of $Z_2 \times Z_2'$, 
 and mass spectra at the tree level 
 of Higgs supermultiplets are shown. 
 The first column represents the corresponding Higgs fields 
before the compactification.}
\label{tb:particles3}
\end{table}

\begin{table}
\begin{center}
 \begin{tabular}[tb]{|l|l|c|c|c|}
\hline
 gauge fields &  $(SU(3)_C, SU(2)_L, U(1)_Z, U(1)_X)$ 
 & $U(1)_R$ & $(Z_2, Z_2')$ &
 mass \\
 \hline\hline
 $V^a$ & 
{\small
 $({\bf 8}, {\bf 1} , {\bf 0}, {\bf 0})$,
 $({\bf 1}, {\bf 3} , {\bf 0}, {\bf 0})$,
 $({\bf 1}, {\bf 1} , {\bf 0}, {\bf 0})$,
 $({\bf 1}, {\bf 1} , {\bf 0}, {\bf 0})$
}
 &0  & $(+, +)$ & $\frac{2n}{R}$ \\
 $V^{\hat a}$ & 
{\small
 $({\bf 3}, {\bf 2} , {\bf -5/6}, {\bf 0})$
 $({\bf \overline{3}}, {\bf 2} , {\bf 5/6}, {\bf 0})$
}
 &0  & $(+, -)$ & $\frac{2n+1}{R}$ \\
 $\Sigma^{\hat a}$ & 
{\small
 $({\bf 3}, {\bf 2} , {\bf -5/6}, {\bf 0})$
 $({\bf \overline{3}}, {\bf 2} , {\bf 5/6}, {\bf 0})$
}
 &0  & $(-, +)$ & $\frac{2n+1}{R}$ \\
 $\Sigma^{ a}$ & 
{\small
 $({\bf 8}, {\bf 1} , {\bf 0}, {\bf 0})$,
 $({\bf 1}, {\bf 3} , {\bf 0}, {\bf 0})$,
 $({\bf 1}, {\bf 1} , {\bf 0}, {\bf 0})$,
 $({\bf 1}, {\bf 1} , {\bf 0}, {\bf 0})$
}
 &0  & $(-, -)$ & $\frac{2n+2}{R}$ \\
 \hline
 \end{tabular}
\caption{The gauge quantum numbers 
 after the compactification, 
 the charges of $U(1)_R$ symmetry,
 parity eigenvalues of $Z_2 \times Z_2'$, 
 and mass spectra at the tree level 
 of gauge supermultiplets are shown. }
\end{center}
\label{tb:particles4}
\end{table}

\end{document}